# Impedance-optical Dual-modal Cell Culture Imaging with Learning-based Information Fusion

Zhe Liu, *Student Member, IEEE*, Pierre Bagnaninchi, Yunjie Yang, *Member, IEEE*

*Abstract*—While Electrical Impedance Tomography (EIT) has found many biomedicine applications, a better resolution is needed to provide quantitative analysis for tissue engineering and regenerative medicine. This paper proposes an impedance-optical dual-modal imaging framework, which is mainly aimed at high-quality 3D cell culture imaging and can be extended to other tissue engineering applications. The framework comprises three components, i.e., an impedance-optical dual-modal sensor, the guidance image processing algorithm, and a deep learning model named multi-scale feature cross fusion network (MSFCF-Net) for information fusion. The MSFCF-Net has two inputs, i.e., the EIT measurement and a binary mask image generated by the guidance image processing algorithm, whose input is an RGB microscopic image. The network then effectively fuses the information from the two different imaging modalities and generates the final conductivity image. We assess the performance of the proposed dual-modal framework by numerical simulation and MCF-7 cell imaging experiments. The results show that the proposed method could significantly improve image quality, indicating that impedance-optical joint imaging has the potential to reveal the structural and functional information of tissue-level targets simultaneously.

## I. INTRODUCTION

3D cell culture has far-reaching significance because it can better mimic the function of living tissues compared with cell monolayers, which has a significant impact on drug screening [1][2]. Providing better models of cell behaviors may be beneficial to the research and treatment of human diseases and reduce animal testing. A key challenge in 3D cell culture is to determine the cellular state in depth and across time. Therefore, a suitable imaging technique is desired to monitor 3D cell culture continuously and non-destructively. Electrical Impedance Tomography (EIT) is a tomographic imaging technique that can recover the conductivity distribution within the interior of a domain through boundary current injection and induced voltage measurements [3]-[5]. It is well known that cell viability can be inferred by measuring its cellular electrical parameters [6]. Recently, miniaturized EIT has been introduced to image the conductivity distribution of the 3D cultivated cells in both static and dynamic setups [7]-[10]. However, the low spatial resolution of EIT has become one of the limiting factors to perform quantitative analysis of the properties and behaviors of 3D cultivated cells in such tissue engineering applications.

In the past, efforts to improve EIT image quality have been mainly focused on advancing the image reconstruction algorithm. A prevailing type of image reconstruction methods is based on regularization, which incorporates certain prior knowledge into the procedure. The state-of-the-art regularization methods for EIT image reconstruction include Total Variation (TV) regularization [11]-[13], Fidelity-Embedded Regularization [14], sparse regularization [15][16] and Adaptive Group Sparsity (AGS) regularization [17][18] etc. Some of these algorithms, e.g. AGS, have been demonstrated to be effective to improve image quality by introducing the structural features of imaging objects as prior knowledge. Recently, deep learning [19] demonstrates its effectiveness in Computer Vision (CV), Natural Language Processing (NLP) and inverse problems in imaging. Deep learning based EIT image reconstruction methods have also been in the ascendant. S. J. Hamilton et al. [20] and M. Capps et al. [21] combined the D-Bar method with deep learning to conduct high-quality EIT image reconstruction. Z. Wei et al. [22] proposed a CNN-based method to solve the inverse problem of EIT. These methods usually utilize the intermediate result generated by certain model-based image reconstruction algorithms and further refine it through a dedicated network. In addition, end-to-end deep-learning-based methods are also reported to solve the inverse problem of EIT [23]-[25]. Though these works showed considerable improvement of EIT image quality, they are focused on solving the single-modal imaging problem. Meanwhile, dual-modal or multi-modal methods are also explored to supplement EIT. For instance, the joint imaging of EIT and ultrasound tomography was investigated and showed improved image quality and structure preservation [26][27]. Li et al. integrated structural information from X-ray tomography into EIT inversion by using the cross-gradient method [28]. These work has shown evidence that multi-modal imaging can improve the EIT image quality by combining complementary information.

Inspired by multi-modal imaging and deep learning, and to improve EIT image quality and promote EIT based quantitative cellular assay in tissue engineering, we propose an impedance-optical dual-modal imaging framework to enable dual-modal cell imaging and learning-based dual-modal information fusion. This work focusses on 2D imaging of the 3D cell culture process, which reconstructs the cross-section of the 3D targets. The imaging framework comprises the impedance-optical dual-modal miniature sensor for cell imaging, the guidance image processing algorithm for optical image preprocessing and a dual-input deep learning model for information fusion and image reconstruction. The advantages of the proposed approach are:

1) Compared with single-modal methods, e.g., regularization and learning-based methods, the proposed framework can generate an EIT image with more accurate target shapes by introducing optical imaging, thereby leading to more precise conductivity distribution estimation.
2) The framework develops a new indirect information fusion approach that addresses the challenge of directly using the optical image to train the deep learning model by converting it to a binary mask image. This approach can be extended to other learning-based multi-modal image reconstruction scenarios with similar issues.

The remainder of the paper is organized as follows. Section II states the principle of EIT inverse problem. Section III describes the proposed framework. Section IV elaborates 2D simulation data generation and experiment setup. Section V illustrates simulation and experimental results. Finally, Section VI draws conclusion and discusses future work.

## II. INVERSE PROBLEM OF EIT

We describe the principle of EIT image reconstruction based on the 16-electrode configuration, as in this work, we adopt a 16-electrode miniature sensor and conduct 2D imaging. Suppose the sensing area occupies a 2D circular region $\Omega \subset \mathbb{R}^2$ (see Fig. 1). Sixteen electrodes denoted by $(e_1, e_2, \ldots, e_{16})$ are attached around the boundary $\partial\Omega$ (see Fig. 1). Adjacent protocol [29] is adopted, where a current $J$ is applied successively to the electrode pairs $(e_\ell, e_{\ell+1}), \ell = 1, \ldots, 16, e_{16+1} := e_1$; and the voltage difference between all other pairs of neighboring electrodes are collected. Let $\sigma = \sigma(x)$ and $u = u(x)$ denote the conductivity distribution and the electrical potential distribution in $\Omega$ respectively, the forward problem of EIT based on the Complete Electrode Model (CEM) [30] can be expressed as:

$$\nabla \cdot (\sigma(x)\nabla u(x)) = 0, \quad x \in \Omega \qquad (1)$$

$$u(x) + z_l \sigma(x)\frac{\partial u(x)}{\partial n} = U_\ell, \quad x \in e_\ell, \ell = 1,2,\ldots,16 \qquad (2)$$

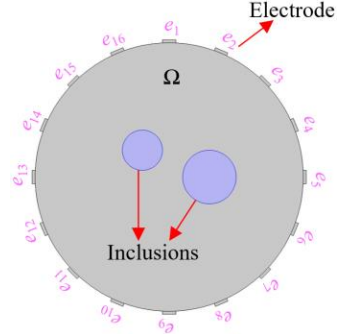

Fig. 1. Sixteen-electrode circular EIT sensor with two inclusions in it.

$$\int_{e_\ell} \sigma(x)\frac{\partial u(x)}{\partial n} dS = J_\ell, \quad \ell = 1,2,\ldots,16 \qquad (3)$$

$$\sigma \frac{\partial u(x)}{\partial n} = 0, \quad x \in \partial\Omega \setminus \bigcup_{l=1}^{16} e_\ell \qquad (4)$$

where $n$ is outward unit normal of $\partial\Omega$. $U_\ell$ and $I_\ell$ denotes respectively the electrical potential and injected current on electrode $e_\ell$.

The existence and uniqueness of the solution $u$ should also be ensured by the charge conservation and the choice of the ground voltage defined respectively by the left and right equations below.

$$\sum_{\ell=1}^{16} J_\ell = 0, \qquad \sum_{\ell=1}^{16} U_\ell = 0 \qquad (5)$$

We define the measured voltage difference between electrode pairs $(e_g, e_{g+1}), g = 1,2,\ldots 16, e_{16+1} := e_1$, subject to the $\ell^{th}$ current injection as:

$$V^{\ell,g} := U_g^\ell - U_{g+1}^\ell \qquad (6)$$

where $U_g^\ell$ and $U_{g+1}^\ell$ denote respectively the measured potential on the $g^{th}$ and $(g+1)^{th}$ electrode.

Time-difference EIT (td-EIT) reconstructs the conductivity variation in $\Omega$ through boundary voltage variation measurements. In this work, the electrodes directly contact the highly conductive cell culture media, and the contact impedance is negligible. Therefore, the boundary voltage variation on the $g^{th}$ electrode pair subject to the $\ell^{th}$ injection can be formulated as:

$$V^{\ell,g}_{\sigma_1} - V^{\ell,g}_{\sigma_0} = -\int_\Omega (\sigma_1 - \sigma_0)\nabla u^\ell_{\sigma_1} \nabla u^g_{\sigma_0} dx \qquad (7)$$

where $\sigma_1$ denotes the conductivity distribution at the observation time point and $\sigma_0$ represents the conductivity distribution at the reference time point. $V^{\ell,g}_{\sigma_1}$ is the $V^{\ell,g}$ corresponding to $\sigma_1$, and so does $V^{\ell,g}_{\sigma_0}$. $u^\ell_{\sigma_1}$ is the electrical potential distribution subject to the $\ell^{th}$ injection and $\sigma_1$. The definition of $u_{\sigma_0}^{\mathcal{G}}$ is similar with $u_{\sigma_1}^{\ell}$. Therefore, (7) can be reformulated as:

$$\Delta V^{\ell,\mathcal{G}} = H^{\ell,\mathcal{G}}(\Delta\sigma) \qquad (8)$$

where $\Delta\sigma$ is the conductivity variation in $\Omega$ and $H^{\ell,\mathcal{G}}$ is the non-linear mapping from $\Delta\sigma$ to $\Delta V^{\ell,\mathcal{G}}$. By eliminating repetitive data according to the reciprocity principle [31], we can obtain a frame of independent measurements, i.e. $\Delta V \in \mathbb{R}^{104}$. Therefore, the forward mapping can be ultimately expressed as $\Delta V = [H^{1,3}, ..., H^{1,15}, H^{2,4}, ..., H^{2,16}, ..., H^{14,16}]^T \triangleq H(\Delta\sigma)$, $H$ is the non-linear mapping from $\Delta\sigma$ to $\Delta V$. The inverse problem can be formulated as:

$$\Delta\sigma = H^{-1}(\Delta V) \qquad (9)$$

where $H^{-1}$ is the inverse mapping operator of $H$, which is to be approximated.

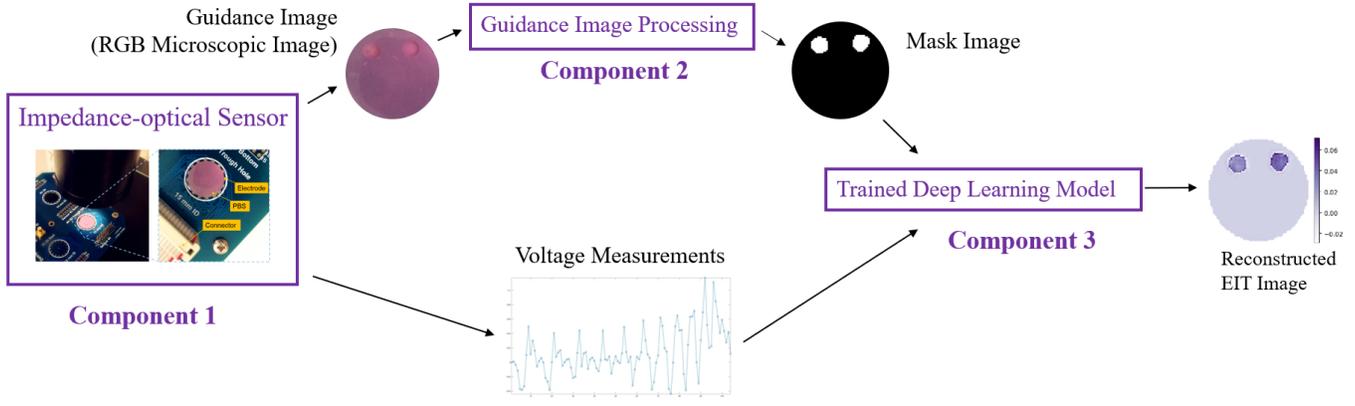

Fig. 2. Schematic of impedance-optical dual-modal imaging framework.

### III. IMPEDANCE-OPTICAL DUAL-MODAL IMAGING FRAMEWORK

In this Section, an impedance-optical dual-modal imaging framework (see Fig. 2) is proposed to improve EIT image quality for 3D cell imaging. It consists of three components, i.e., the impedance-optical miniature sensor, the guidance image processing algorithm, and a deep learning model. First, the impedance-optical sensor will simultaneously output a frame of voltage measurements and an RGB microscopic image named the guidance image ($I^g$). Then, the guidance image processing algorithm will convert $I^g$ into its corresponding mask image ($I^m$). Finally, $I^m$ and the voltage measurements are fed into a deep learning model to generate the reconstructed EIT image.

#### A. Impedance-optical Dual-modal Sensor

The dual-modal sensor (see Fig. 3) combines a miniature 16-electrode EIT sensor with a digital microscope (Digital USB Microscope 1.3M, RS Components Ltd). The EIT sensor is manufactured on a Printed Circuit Board (PCB). A transparent glass substrate is attached at the bottom of the sensing area to support cells and enable optical imaging. The height and diameter of the sensing chamber are 1.6 mm and 14 mm, respectively. The 16 gilded microelectrodes are manufactured using the half-hole process and distributed at the periphery of the sensing area. The digital microscope is placed over the sensing chamber and is calibrated well to make its view filed the same as the sensing area. This dual-modal sensor can then simultaneously record the cells' visual profiles and EIT measurements.

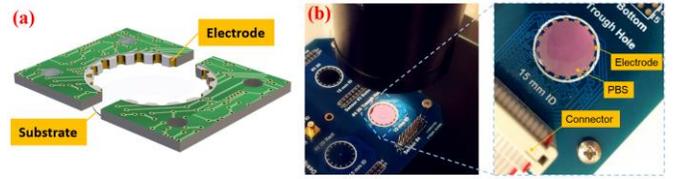

Fig. 3. Impedance-optical dual-modal sensor. (a) EIT sensor structure. (b) The manufactured dual-modal sensor.

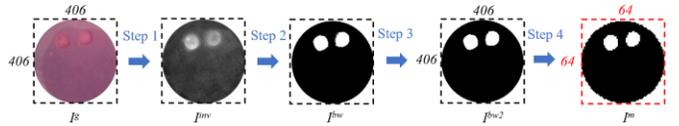

Fig. 4. An illustration of the guidance image processing procedure. The dashed square represents the circumscribed square region of the circular sensing region. The numbers mean the number of pixels for each side of the square.

#### B. Guidance Image Processing

Guidance image processing containing four steps converts the guidance image $I^g$ into its corresponding mask image $I^m$ (see Fig. 4). The size of $I^m$ is the same as that of the expected EIT image, which occupies a circular region inscribed in a $64 \times 64$ square region, while the size of $I^g$ is much larger than it. $I^g$ also occupies a circular region, but this circle inscribes in a $406 \times 406$ square region. It should also be noted that $I^g$ has three color channels, i.e., R, G and B. Therefore, this algorithm starts with the processing of the high-resolution RGB image $I^g$.

In $I^g$, the illumination often causes shadow, which is invalid information and will significantly affect the target segmentation. Besides, as the structure of the targets is only desired, preservation of color has seldom significance. Therefore, the first step is to obtain the 1D illuminant invariant image $I^{inv}$ of $I^g$ following the methods proposed by Finlayson et al. [32] in order to convert $I^g$ into a grey-scale image while removing the influence of illumination. The equation is formulated as:

$$I^{inv}(r,c) = \exp(\chi_1(r,c)\cos(\Theta) + \chi_2(r,c)\sin(\Theta)) \quad (10)$$

where $r$ and $c$ are pixel indexes. $\Theta$ is the projection direction in the 2D log-chromaticity space of $I^g$ which is a constant for a specific camera. This direction leads to the minimum Shannon's entropy for $I^{inv}$ and can be approximately obtained by traversing every integer angle from $1°$ to $180°$. $\chi_1(r,c)$ and $\chi_2(r,c)$ is expressed as:

$$[\chi_1(r,c), \chi_2(r,c)]^T = U\rho(r,c) \quad (11)$$

Here, $U$ is a $2 \times 3$ orthogonal matrix and take the value of $U = [v_1, v_2]^T$, $v_1 = \left[\frac{1}{\sqrt{2}}, -\frac{1}{\sqrt{2}}, 0\right]^T$, $v_1 = \left[\frac{1}{\sqrt{6}}, \frac{1}{\sqrt{6}}, -\frac{2}{\sqrt{6}}\right]^T$. $\rho(r,c)$ is defined by:

$$\rho(r,c) = \left[\ln\left(\frac{R(r,c)}{\Xi(r,c)}\right), \ln\left(\frac{G(r,c)}{\Xi(r,c)}\right), \ln\left(\frac{B(r,c)}{\Xi(r,c)}\right)\right]^T \quad (12)$$

where, $\Xi(r,c) = \sqrt[3]{R(r,c)G(r,c)B(r,c)}$. $R(r,c)$, $G(r,c)$ and $B(r,c)$ are the three components of a color image.

Then, the binary version of $I^{inv}$ can be generated by using the following thresholding segmentation method:

$$I^{bw}(r,c) = \begin{cases} 0, & \text{if } I^{inv}(r,c) < \beta \\ 1, & \text{if } I^{inv}(r,c) \geq \beta \end{cases} \quad (13)$$

where $I^{bw}$ denotes the binary image after thresholding. The threshold value $\beta$ is selected based on empirical trials. However, such method may lead to irregular boundary and randomly distributed white pixels. To address this issue, the third step applies morphological operations to $I^{bw}$ to acquire a clean binary image with boundary-regular targets. In this paper, open operation (14) and dilation operation (15) are successively applied to reduce background irrelevant information and recover accurate target profiles. The two operations are defined as [33]:

$$I^{bw1} = I^{bw} \circ S = \cup\{(S)_z | (S)_z \subseteq I^{bw}\} \quad (14)$$

$$I^{bw2} = I^{bw1} \oplus S = \left\{z \middle| \left(\hat{S}\right)_z \cap I^{bw1} \neq \emptyset\right\} \quad (15)$$

where, $I^{bw} \circ S$ means $I^{bw}$ is opened by the structuring element $S$ and $I^{bw1} \oplus S$ means $I^{bw1}$ is dilated by $S$. $(S)_z$ and $\hat{S}$ are defined as [33]:

$$(S)_z = \{k | k = a + z, a \in S\} \quad (16)$$

$$\hat{S} = \{w | w = -a, a \in S\} \quad (17)$$

where $z = (z_1, z_2)$ is a fixed point in the image space where $I^{bw}$ and $I^{bw1}$ exist.

$I^{bw2}$ already provides the expected structural information, but it cannot be directly used as the input of the MSFCF-Net. As stated at the beginning of this subsection, the size of $I^m$ is required to be the same as that of the EIT image. In addition, the height and width of images generated by the first three steps (i.e., $I^{inv}$, $I^{bw}$ and $I^{bw2}$) is the same as those of $I^g$. Therefore, the final step is to down-sample $I^{bw2}$ into a smaller circular image internally tangent with a $64 \times 64$ square region. The resulting smaller image is the $I^m$, and it is exactly the tiny version of $I^{bw2}$.

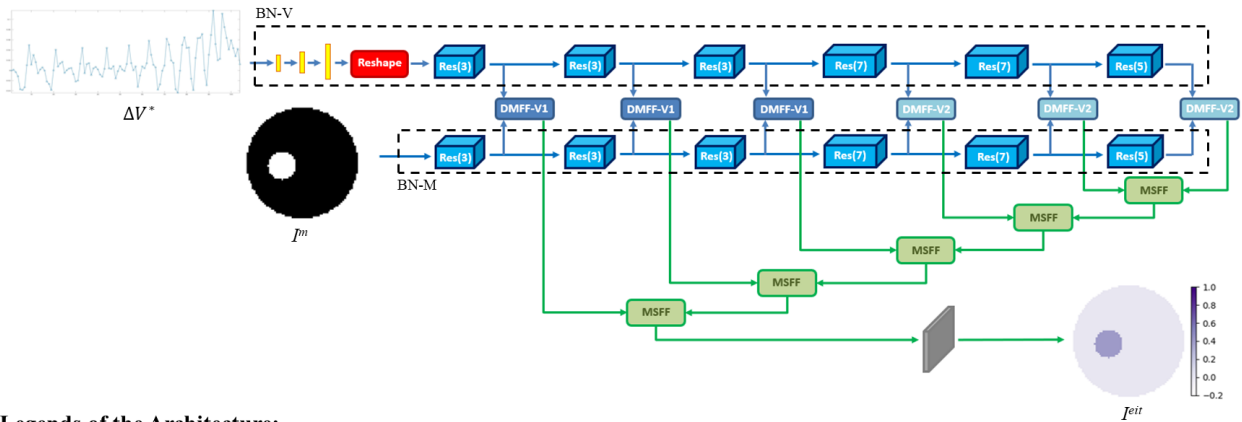

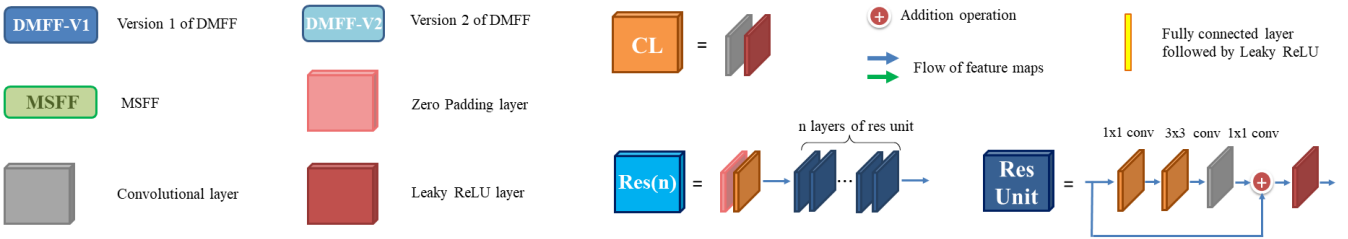

Fig. 5 Architecture of MSFCF-Net. Note, the color of arrow is only for indicating the feature maps flowing to different function block. BN-V and BN-M are the layers in dashed black squares.

## C. Multi-scale Feature Cross Fusion Network

MSFCF-Net reconstructs an EIT image $I^{eit}$ from a frame of voltage measurements $\Delta V^* \in \mathbb{R}^{104}$ and a mask image $I^m$. We describe $I^{eit}$ and $I^m$ with a tensor of size $C \times 64 \times 64$, where $C = 1$ denotes the number of channels for a multi-channel image. $\Delta V^*$ and $I^{eit}$ are defined by:

$$\Delta V^* = \frac{V_{\sigma_1} - V_{\sigma_0}}{V_{\sigma_0}} \tag{18}$$

$$I^{eit} = -\frac{\sigma_1 - \sigma_0}{\sigma_0} \tag{19}$$

As the definition stated in section II, td-EIT aims to recover $\Delta\sigma = \sigma_1 - \sigma_0$ from $\Delta V = V_{\sigma_1} - V_{\sigma_0}$. However, we adopt the relative changes to help facilitate the training of a deep learning model [34].

Our goal is to learn an end-to-end mapping $F$ from $\Delta V_i^*$ and $I^m$ to $I^{eit}$. Given a training dataset $\{\Delta V_i^*, I_i^m, I_i^{eit}\}_{i=1}^N$, the problem can be described as:

$$\hat{\theta} = \arg\min_\theta \frac{1}{N}\sum_{i=1}^N L(F_\theta(\Delta V_i^*, I_i^m), I_i^{eit}) + \frac{\lambda}{2}\|\theta\|_2^2 \tag{20}$$

where the second term is $l_2$ regularization with a penalty parameter $\lambda$, which can reduce over-fitting. $\theta = \{W, b\}$ represents the weights and bias of MSFCF-Net. $L$ is the loss function to minimize the difference between $I_i^{eit}$ and $F_\theta(\Delta V_i^*, I_i^m)$. As EIT image reconstruction is a regression problem, the mean squared error loss function is used, and $L$ is defined as:

$$L(F_\theta(\Delta V_i^*, I_i^m), I_i^{eit}) = \|F_\theta(\Delta V_i^*, I_i^m) - I_i^{eit}\|_2^2 \tag{21}$$

The architecture of MSFCF-Net is shown in Fig. 5. Subnetworks in MSFCF-Net can be divided into three categories, i.e. the backbone networks, dual-modal feature fusion modules, and multi-scale feature fusion modules.

### 1) Backbone Networks (BN)

The backbone network extracts latent features from inputs. Thus, this network should have a good ability of feature extraction. The Darknet as the backbone of YOLOV3 is proved effective and powerful on feature extraction [35]. Inspired by its architecture, we designed the Darknet-like backbone networks for our application. The backbone network for voltage measurements (BN-V) has three additional fully connected layers followed by a reshape operation because of the dimension difference between $Q$ and $I^m$ (see Fig. 5). The output of the reshape operation is a feature map with the size of $1 \times 64 \times 64$. The rest of BN-V is the same as the backbone network for mask image (BN-M), which consists of five residual blocks denoted by Res(n). Res(n) starts with left-and-upper zero padding followed by a $Conv + Leaky\ ReLU$ unit with $Kernel\ Size = 3 \times 3$, $Stride\ Step = 2$, and the number of kernels is twice as that of input feature maps. Then $n$ residual units (represented by Res Unit, see Fig. 2) follows. The idea of Res Unit is proposed in [36], in which the short connection can make the deep network easier to train. Therefore, the combination of mentioned components in Res(n) will make the height and width of the output feature maps half than those of input feature maps while the number of feature maps doubles.

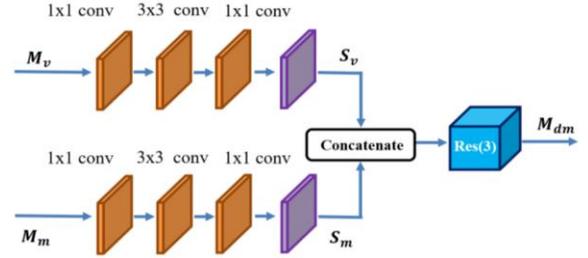

Fig. 6. Architecture of DMFF. The purple block means the mapping $f^A$. $f^A$ equals to $f^{CA}$ for DMFFM-V1 and $f^A$ equals to $f^{SA} \circ f^{CA}$ for DMFFM-V2. The meaning of other components is the same as those of legends in Fig. 5.

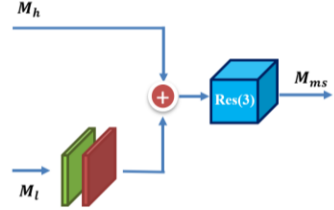

Fig. 7. Architecture of MSFF. The green block means transposed convolution layer. The meaning of other components is the same as those of legends in Fig. 5.

### 2) Dual-modal Feature Fusion Module (DMFF)

Dual-modal Feature Fusion Modules (DMFF) fuse information from different sources (see Fig. 6). To maintain the main information and eliminate the trivial ones, the attention mechanism which is originally used in natural language processing [37] is adopted in DMFF. As the feature maps generated by each layer in CNN have both channel dimension and spatial dimension, there are two types of attention mechanism, i.e., the channel-wise attention and spatial-wise attention. In BN-V and BN-M, with the increase of the number of layers, the spatial dimension gradually decreases. For the feature maps with a small spatial dimension, the spatial information is totally lost, and information carried by this type of feature maps is usually called semantic information. The spatial relationship between each element of the feature maps is trivial. Therefore, there are two types of DMFF in MSFCF-Net, i.e., DMFF-V1 and DMFF-V2 (see Fig. 5). DMFF-V1 corresponds to the feature maps with large spatial dimension, and it will incorporate both channel-wise attention and spatial-wise attention. DMFF-V2 corresponds to the feature maps with small spatial dimension, and it will only incorporate the channel-wise attention. The implementation of attention mechanisms adopts the convolutional block attention modules proposed in [38], which is proved to be an effective and efficient method. Suppose the mapping of channel attention module in CBAM is denoted by $f^{CA}$ and that of spatial attention module is denoted by $f^{SA}$, the mappings of both DMFF-V1 and DMFF-V2 can be uniformly expressed as:

$$S_v = f^A\left(f^{CL}_{1,1}\left(f^{CL}_{3,1}\left(f^{CL}_{1,1}(M_v)\right)\right)\right) \quad (22)$$

$$S_m = f^A\left(f^{CL}_{1,1}\left(f^{CL}_{3,1}\left(f^{CL}_{1,1}(M_m)\right)\right)\right) \quad (23)$$

$$M_{dm} = R^3([S_v, S_m]) \quad (24)$$

where $M_v$ is the feature map from BN-V and $M_m$ is the feature map from BN-M. The size of the feature maps $S_v, S_m, M_{dm}, M_v$ and $M_m$ is the same. $f^A$ equals to $f^{CA}$ for DMFFM-V1 and equals to $f^{SA} \circ f^{CA}$ for DMFFM-V2, which is the only difference between the two modules. $f^{CL}_{\cdot,\cdot}$ denotes the mapping for *Conv + Leaky ReLU* unit. The first subscript means the kernel size and the second means the convolution step used in the convolution layer in this unit. $[\cdot,\cdot]$ denotes the concatenation operation and $R^3$ means the mapping of Res(3).

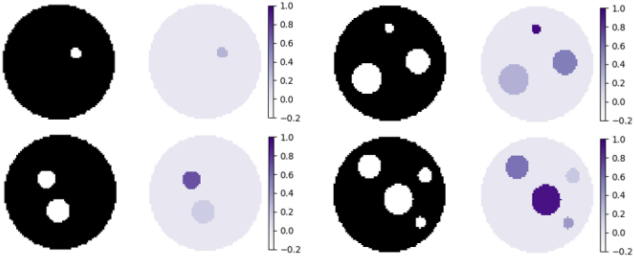

Fig. 8. Examples of simulated EIT images and corresponding mask images. For each pair, the left one is the binary mask image and the right one is the reconstructed EIT image.

### 3) Multi-scale Feature Fusion Module (MSFF)

Feature maps of different scales will provide information of different scales. It will generate a more precise result if information of different scales can be integrated together. Many work in computer vision and image processing demonstrates that fusing feature maps of different scales is an efficient way to improve the performance of the network [35][39][40]. In addition, Chen et al. [24] and Li et al. [23] both adopted this method in their work on EIT image reconstruction and showed good results. MSFCF-Net also adopts the same idea and multi-scale feature fusion module (MSFF, see Fig. 7) undertakes this function. MSFF module uses a simple way to perform information fusion. First, the spatial dimension of low scale feature maps will be enlarged twice by transposed convolutional layers followed by the *Leaky ReLU* layer. Then, the output of *Leaky ReLU* layer and the output of the network block before the current block will be added together in MSFF. The addition operation here is inspired by the work on human eye-fixation prediction, where the authors also face a dual-modal information fusion problem and fuse information of different scales by addition operation [41]. Like YOLOV3, the initial fused feature maps will be fed into multiple layers for thoroughly information fusion. Instead of using successive convolutional layers, the basic module in BN, i.e. Res(n), is used in MSFF to conduct post information fusion. Because this module has a satisfactory feature extraction ability while it can prevent the degradation of the network [36]. Finally, the output of current MSFF will be the input of the next MSFF. The mapping of MSFF can be represented as:

$$M_{ms} = R^3\left(M_h + f^{TCL}_{2,2}(M_l)\right) \quad (25)$$

where, $M_l$ is the low-scale feature map and $M_h$ is the high-scale feature map. The size of the output feature map $M_{ms}$ is the same with that of $M_h$. $f^{TCL}_{2,2}$ represents the mapping for *Transposed Conv + Leaky ReLU* unit. The first subscript means the kernel size and the second represents the convolutional step in the convolution layer of this unit. $R^3$ means the mapping of Res(3).

TABLE I
QUANTITATIVE METRICS FOR COMPARING DIFFERENT ALGORITHMS ON TEST SET

| Algorithms | Noise-free | | 50dB | | 40dB | | 30dB | |
|---|---|---|---|---|---|---|---|---|
| | M-RIE | M-MSSIM | M-RIE | M-MSSIM | M-RIE | M-MSSIM | M-RIE | M-MSSIM |
| Treg-GL | 1.1150 | 0.3663 | 1.1151 | 0.3650 | 1.1155 | 0.3570 | 1.1235 | 0.3090 |
| SBL | 1.8382 | 0.7227 | 1.8355 | 0.7214 | 1.8115 | 0.7110 | 1.7279 | 0.6312 |
| CG | 1.0470 | 0.4028 | 1.0470 | 0.4022 | 1.0471 | 0.3981 | 1.0488 | 0.3694 |
| FC-UNet | 0.6584 | 0.8295 | 0.6582 | 0.8295 | 0.6587 | 0.8292 | 0.6595 | 0.8271 |
| S-MSFCF-Net | 0.6230 | 0.8418 | 0.6230 | 0.8418 | 0.6230 | 0.8417 | 0.6270 | 0.8403 |
| **MSFCF-Net** | **0.4388** | **0.9460** | **0.4388** | **0.9460** | **0.4388** | **0.9460** | **0.4386** | **0.9460** |

TABLE II
COMPARISON OF DIFFERENT ALGORITHMS ON DIFFERENT TYPES OF SAMPLES

| Algorithms | 1-Object | | 2-Object | | 3-Object | | 4-Object | |
|---|---|---|---|---|---|---|---|---|
| | M-RIE | M-MSSIM | M-RIE | M-MSSIM | M-RIE | M-MSSIM | M-RIE | M-MSSIM |
| FC-UNet | 0.4770 | 0.9510 | 0.6558 | 0.8717 | 0.7327 | 0.7884 | 0.7559 | 0.7156 |
| S-MSFCF-Net | 0.4024 | 0.9622 | 0.6176 | 0.8826 | 0.6918 | 0.8072 | 0.7662 | 0.7237 |
| **MSFCF-Net** | **0.1305** | **0.9988** | **0.4738** | **0.9559** | **0.5651** | **0.9273** | **0.5674** | **0.9054** |

TABLE III
VISUAL COMPARISON OF DIFFERENT ALGORITHMS ON FIVE REPRESENTATIVE SAMPLES (LEFT COLUMN: RECONSTRUCTION; RIGHT COLUMN: ERROR IMAGE)

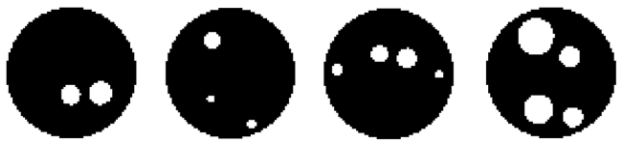

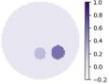

Fig. 9. Mask images corresponding to samples in Table III.

## IV. DATA GENERATION AND EXPERIMENTAL SETUP

### A. Sensor Modelling and Dataset Generation

We establish the training, validation and test sets to train and evaluate the proposed MSFCF-Net. We modelled the 2D 16-electrode circular EIT sensor in COMSOL Multiphysics and solve the forward problem of EIT to generate simulation data.

To make the deep learning model suitable for 3D cell culture imaging, specifically for 3D cell spheroid imaging, we considered multi-level, multi-circular-object conductivity distributions. In the sensing area, we generate four types of data and a sample belonging to a certain type of data includes a fixed number of objects (from one to four). For a certain type of data, for example, the one including three objects, we assigned three non-overlapping circular objects with random diameters (from 0.03d to 0.3d, d is the diameter of the sensing area), positions, and random conductivity values (from $0.0001\ S \cdot m^{-1}$ to $0.05\ S \cdot m^{-1}$). The background conductivity is $0.05\ S \cdot m^{-1}$. Mask images for training are also generated in simulation by a simple approach of assigning number one to pixels where there are objects and assigning number zero to rest pixels. Four examples of simulated conductivity images and corresponding binary mask images are illustrated in Fig. 8.

Based on the settings stated above, we built a dataset of 29,333 samples and each sample includes a frame of voltage measurements, the true conductivity image, and a mask image. There are 7,035 1-object samples, 7,298 2-object samples, 7,500 3-object samples and 7,500 4-object samples. In order to maintain the data balance in training and evaluation, for each type of data, we randomly select 10% samples as the test set and select 10% samples from the rest data as the validation set. The rest will serve as the training set. Eventually, we have 23,762 samples for training, 2,639 samples for validation and 2,932 samples for testing.

### B. Dual-modal Imaging System Setup

The dual-modal sensor is connected to the in-house developed EIT system [42] to collect real-world experimental data. The frequency of the injected current is 10kHz. The view field of the digital microscope and the sensing area of the impedance sensor coincide precisely.

### C. Network Training

The MSFCF-Net is implemented using Pytorch, trained and tested on a workstation with a GeForce RTX 2070 Super. AdamW [43] is employed for optimization. The training process is divided into two phases. In the first phase, we select 1-object and 2-object samples to train the network to find appropriate initial parameters for the second phase, which will help with the reduction of training epochs when we fit a large amount of training data in the second phase. In this case, there are 11,612 samples for training and 1,289 samples for validation. The hyper parameters are set as follows: the learning rate is 0.001 and the penalty parameter $\lambda$ of $l_2$ regularization is set as 0.0001. The total number of training epoch is 100 and the batch size of each update is 200.

In the second phase, we use the whole training set (23,762 samples) and the whole validation set (2,639) to train MSFCF-Net. To improve the robustness of our model, additive noise is added to the voltage measurements during the training process. For each type of data, we separately add noise with the Signal-to-Noise Ratio (SNR) of 50dB, 40dB and 30dB to one fourth of the training and validation data, respectively. The hyper parameters of the second training phase are set as follows: the learning rate is 0.0005 and the penalty parameter $\lambda$ is set as 0.00005. The total number of training epoch is 100 and the batch size of each update is 120. The tolerance is set as 10 epochs for early stopping. Finally, the training process is stopped at epoch 93 at the second training phase.

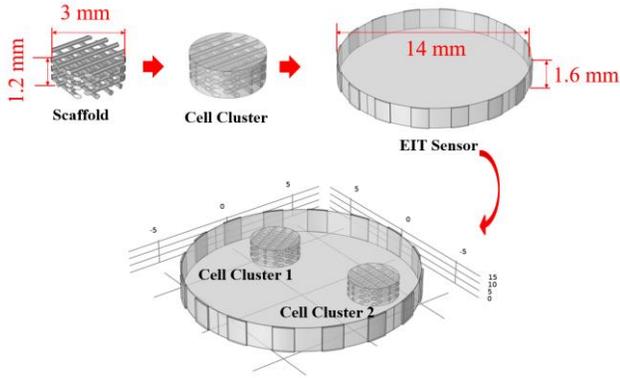

Fig. 10. Modelling of quasi-2D EIT sensor equipped with two scaffolds.

## V. RESULTS AND DISCUSSION

The proposed method is evaluated by numerical simulation and MCF-7 cell spheroids experiments. The performance of MSFCF-Net is compared with other widely used single-modal based EIT image reconstruction algorithms, i.e., Gaussian-Laplace regularization (TReg-GL) [44] and Sparse Bayesian Learning (SBL) [45], and a dual-modal based image reconstruction algorithm using cross-gradient regularization (CG) [28]. In this work, the mask image is used to replace the CT image in [28] as the assisted image in both simulation and real experiments. We also compare with the recently proposed end-to-end deep learning model FC-UNet [24] and the single-modal version of MSFCF-Net (named S-MSFCF-Net). FC-UNet is originally designed for pixel-level classification for EIT image. As we deal with the conductivity distribution prediction as a regression problem, to make the fair comparison, we remove the activation function in the output layer of the FC-UNet and train it with the same loss function and settings as MSFCF-Net. For S-MSFCF-Net, we remove the BN-M and DMFF modules while the MSFF modules will fuse different scales of feature maps from BN-V. S-MSFCF-Net is also trained with the same loss function and settings as MSFCF-Net. It should be emphasized that all reconstructed EIT images either based on numerical simulation or real-world experiments are quantitative images, which means that the value at each pixel in an EIT image denotes the estimated quantity in (19). Besides, as the range of pixel values in images generated by learning-based algorithms belongs to [-0.2, 1] in numerical simulation (this is carefully checked for each image), for simplicity and better comparison, this type of images in Table III, IV and V uniformly adopts the same color bar, whose maximum value is set as 1 and minimum value is set as 0.2.

### A. Numerical Simulation

Relative Image Error (RIE) and mean structural similarity index (MSSIM) [46] are used to quantitatively evaluate the image quality, which are defined as:

TABLE IV
IMAGE RECONSTRUCTION RESULTS BASED ON PERTURBED MASK IMAGES

| Mask Image | Predicted Image | Error Image |
|---|---|---|
| | RIE = 0.1919 | MSSIM = 0.9693 |
| | RIE = 0.2158 | MSSIM = 0.9704 |
| | RIE = 0.1805 | MSSIM = 0.9377 |
| | RIE = 0.1696 | MSSIM = 0.9034 |

$$\text{RIE} = \frac{\|A - B\|_2}{\|B\|_2} \quad (26)$$

$$\text{MSSIM} = \frac{1}{wh}\sum_r \sum_c \text{SSIM}(r, c) \quad (27)$$

where $A$ is the image to be evaluated and $B$ is the selected reference image. $r$ and $c$ are the position indexes of an image. $w$ and $h$ are the weight and height of an image, respectively. $\text{SSIM}(r, c)$ is the structural similarity index map [46], and is defined as:

$$\text{SSIM}(r, c) = \frac{(2\mu_A \mu_B + C_1)(2\delta_{AB} + C_2)}{(\mu_A^2 + \mu_B^2 + C_1)(\delta_A^2 + \delta_B^2 + C_2)} \quad (28)$$

where $\mu_A$, $\mu_B$, $\delta_A$, $\delta_B$, and $\delta_{AB}$ are the local means, standard deviations and cross-covariance for image $A$ and $B$, which are also calculated following methods in [46]. $C_1 = (K_1 L)^2$ and $C_2 = (K_2 L)^2$. $K_1$ and $K_2$ are constants whose values are 0.01 and 0.03, respectively. As the range for reconstructed EIT images in this work is [0, 1], $L$ is set as 1.

During the evaluation process, we calculate RIE and MSSIM for each image in the test set and then average all values. Another two numerical metrics that evaluate performance on the whole dataset level are the mean RIE (M-RIE) and the mean MSSIM (M-MSSIM).

TABLE V
IMAGE RECONSTRUCTION RESULTS FOR QUASI-2D EIT SENSOR

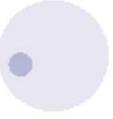

Table I illustrates the quantitative evaluation results at different SNR levels on the test set. It is obvious that the metrics of MSFCF-Net are superior to other given algorithms, indicating the robustness and effectiveness of the proposed dual-modal framework. Deep learning based methods all show better noise-resistance capability than conventional model-based algorithms and their M-RIE and M-MSSIM are at the similar level with the decrease of SNR. Tables II compares the metrics of deep learning models on different types of samples with the SNR=50dB. As reconstructing multi-object and multi-level conductivity distribution is much more challenging, the metrics on 2-object samples have a big drop than those on 1-object samples for all deep learning models. For a specific type of samples, it is evident that the performance of MSFCF-Net is much better than the other two. Especially, though S-MSFCF-Net only removes the mask image related structures from MSFCF-Net, it still cannot reach the performance of MSFCF-Net. The reason is that: single-modal deep learning models will take the duty on both position and structure prediction and conductivity value prediction. But the proposed dual modal deep learning model in essence utilizes more structural information, thus better conductivity prediction can be expected.

Table III shows comparison of five representative phantoms, which are reconstructed from test data with SNR=50dB. GT denotes the ground truth image. Images generated by learning-based algorithms in each row share the same color map (in the rightest column). The left column under each algorithm is the reconstructed EIT image and the right one is the error image which is the absolute difference between the reconstruction and the ground truth. Mask images (from left to right) corresponding to samples in Table III (from top to bottom) for MSFCF-Net are illustrated in Fig. 9. Although both TReg-GL and SBL can predict the position of objects but the shape and conductivity values are always inaccurate (see their error images, RIE and MSSIM). For CG, the reconstructed images are very similar to images by Treg-GL and the quality of images is not improved noticeably according to their error images and numerical metrics. However, if the image generated by CG is zoomed, it is obvious that clear boundaries of objects is visible, which is exactly the results of introducing Cross-Gradient regularization. Thus, the Cross-Gradient regularization can only augment the object boundaries based on the assisted image whilst it cannot essentially improve the EIT image quality. For deep learning based approaches, FC-UNet and S-MSFCF-Net can generate more accurate position, shape and conductivity values, but the errors are still significant. Only MSFCF-Net can reconstruct the best EIT images among the given algorithms with the most accurate position, shape and conductivity values.

In practical applications of the proposed method, inaccurate mask images may be generated due to many factors, such as unideal guidance image processing algorithm or noisy guidance image. To assess the robustness of the proposed method when encountering an inaccurate mask image, Table IV selects the first and fourth samples in Table I for further analysis. Two different random perturbations are applied to the mask images of each sample, which are shown in the first column of the table. Each result occupying one row contains three images, i.e. the input mask image, the predicted EIT image and the error image from left to right. Observing the error images, it is obvious that the conductivity value can still be predicted accurately except for the pixels on the boundary while losing some structural information. Compared with the results generated by MSFCF-Net in Table I, although the quality of the image based on perturbed mask image is lower than the quality of that based on the accurate mask (see RIE and MSSIM), the quality of these images is still much better than the quality of images generated by the model-based algorithms. This analysis implies that, in real-world experiments, we can acquire a quantitatively meaningful EIT image even if the guidance image processing algorithm cannot generate a very accurate mask image.

In many tissue engineering applications, cells are cultured within the scaffold and monitoring of cell growth is vital to the process [8]. Cell growth at different stages within the scaffold will lead to a decrease of conductivity of various levels, which can be mapped by EIT [9]. To further demonstrate the effectiveness of the proposed framework, we simulated the imaging of cell growth within bio-scaffolds [8][9] by using EIT. We modelled a regular-shape scaffold, a quasi-2D EIT sensor with 16 electrodes and a cell culture model with two cell clusters (see Fig. 10). The modelled sensor has the same dimension as the real sensor in Fig. 3. The height and diameter of the scaffold are 1.2 mm and 3 mm, respectively. The background conductivity is set as 0.05 S/m and the conductivity of the scaffold material is set as $10^{-8}$ S/m. The cells are modelled as evenly distributed in the space among scaffolds. We modelled two scenarios. The first contains one cell cluster and the conductivity of the cells cluster is set as 0.025 S/m. The second has two cell clusters to simulate cell growth at two different stages. The conductivities in cell cluster 1 and cell cluster 2 are set as 0.02 S/m and 0.04 S/m, respectively. In these cases, the reference conductivity distribution for the first scenario is the homogeneous medium whose conductivity is 0.05 S/m with a scaffold and the reference conductivity distribution for the second scenario is the same homogeneous medium including two scaffolds. Therefore, only the cell's conductivity contributes to the predicted conductivity variation, which is also indicated in [8].

TABLE VI
COMPARISON BETWEEN DIFFERENT ALGORITHMS ON MCF-7 CELL EXPERIMENTS

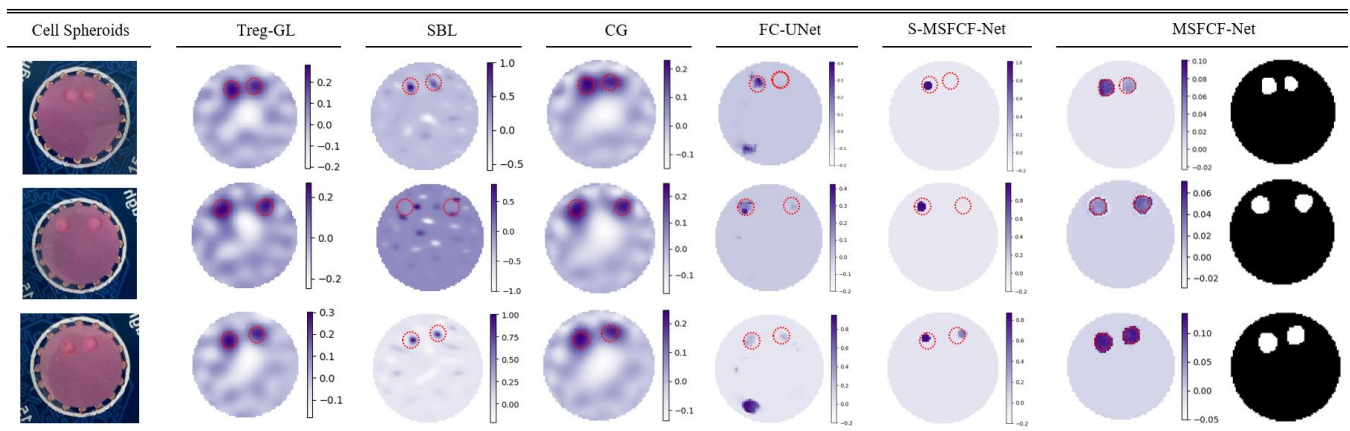

Table V gives the image reconstruction results under settings described in last paragraph. By using the proposed approach, we could obtain reconstruction images with RIE lower than 0.23 and SSIM larger than 0.97. The results show strong evidence that the proposed framework can generate accurate conductivity distribution under a different setting. It also presents good generalization ability when dealing with the challenging scenario of scaffold-based cell culturing imaging.

### B. Cell Experiments

The performance of the proposed framework is further evaluated on data collected from real-world experiments (see Table VI). The imaging target is MCF-7 cell spheroids (diameter ∼ 2 mm). The rightest column is the mask image generated from the guidance image processing algorithms stated in Subsection B, Section III. The threshold values $\beta$ in (13) for the three guidance images (from top to bottom) are set as 0.66, 0.45 and 0.5 respectively based on a series of trials. The $3 \times 3$ kernel is adopted in the next two morphological operations for all cases.

In the Table VI, red dash line denotes the location of the cell spheroids. For conventional single-modal based and dual-modal based algorithms, the cell spheroid structure is totally lost and the reconstructed images contain too much unmeaningful information in the background, although they can locate the position of cell spheroids. As discussed in Subsection A, Section V, the augmented object boundary is visible in images reconstructed by CG whilst the quality of these images is not essentially improved. All the deep learning models can eliminate irrelevant information in the background. Moreover, the proposed MSFCF-Net reconstructs EIT images with the best image quality compared with the other learning-based methods.

## VI. CONCLUSION AND FUTURE WORK

In this paper, we proposed an impedance-optical dual-modal imaging framework for 3D cell culture imaging. We combined optical imaging with EIT to tackle the low image quality issue of EIT. We also developed a learning-based approach to fuse the dual-modal information and reconstruct high-quality conductivity images. The results on simulation data and real-world data on MCF-7 cell spheroids demonstrate that the proposed framework could generate a more accurate estimation of conductivity distribution, which implies the possibility of quantitative imaging for EIT in tissue engineering. Future research will deal with the situation when the structure of the object in the mask image suffers more severe perturbation and develop more advanced image processing algorithms to generalize the method to other optical imaging approaches for tissue engineering, e.g. optical coherence tomography.

## REFERENCES

[1] F. Pampaloni et al., "The third dimension bridges the gap between cell culture and live tissue," in *Nature Reviews Molecular Cell Biology*, vol. 8, no. 10, pp. 839-845, 2007.
[2] E. Cukierman et al., "Taking cell-matrix adhesions to the third dimension," in *Science*, vol. 294, no. 5547, pp. 1708-1712, 2001.
[3] R. H. Bayford, "Bioimpedance tomography (electrical impedance tomography)," in *Annu. Rev. Biomed. Eng.*, vol. 8, pp. 63-91, 2006
[4] L. Miao, Y. Ma and J. Wang, "ROI-based image reconstruction of electrical impedance tomography used to detect regional conductivity variation," in *IEEE Transactions on Instrumentation and Measurement*, vol. 63, no. 12, pp. 2903-2910, 2014.
[5] J. Yao et al., "Development of three-dimensional integrated microchannel-electrode system to understand the particles' movement with electrokinetics," in *Biomicrofluidics*, vol. 10, no. 2, p. 024105, 2016.
[6] R. Pethig and D. B. Kell, "The passive electrical properties of biological systems: their significance in physiology, biophysics and biotechnology," in *Physics in Medicine & Biology*, vol. 32, no. 8, p. 933, 1987.
[7] Y. Yang, J. Jia, S. Smith, N. Jamil, W. Gamal, and P. O. Bagnaninchi, "A miniature electrical impedance tomography sensor and 3-D image reconstruction for cell imaging," in *IEEE Sensors Journal*, vol. 17, no. 2, pp. 514-523, 2016.
[8] Y. Yang, H. Wu, J. Jia, and P. O. Bagnaninchi, "Scaffold-based 3-d cell culture imaging using a miniature electrical impedance tomography sensor," in *IEEE Sensors Journal*, vol. 19, no. 20, pp. 9071-9080, 2019.

[9] H. Wu, W. Zhou, Y. Yang, J. Jia, and P. O. Bagnaninchi, "Exploring the potential of electrical impedance tomography for tissue engineering applications," in *Materials*, vol. 11, no. 6, p. 930, 2018.

[10] H. Wu, Y. Yang, P. O. Bagnaninchi and J. Jia, "Electrical impedance tomography for real-time and label-free cellular viability assays of 3D tumour spheroids," in *Analyst*, vol. 143, no. 17, pp. 4189-4198, 2018.

[11] J. Liu, L. Lin, W. Zhang, and G. Li, "A novel combined regularization algorithm of total variation and Tikhonov regularization for open electrical impedance tomography," in *Physiological Measurement*, vol. 34, no. 7, p. 823, 2013.

[12] B. Gong *et al.*, "Higher order total variation regularization for EIT reconstruction," in *Medical & Biological Engineering & Computing*, vol. 56, no. 8, pp. 1367-1378, 2018.

[13] A. Borsic, B. M. Graham, A. Adler and W. R. B. Lionheart, "In Vivo Impedance Imaging with Total Variation Regularization," in *IEEE Transactions on Medical Imaging*, vol. 29, no. 1, pp. 44-54, Jan. 2010, doi: 10.1109/TMI.2009.2022540.

[14] K. Lee, E. J. Woo and J. K. Seo, "A fidelity-embedded regularization method for robust electrical impedance tomography," in *IEEE Transactions on Medical Imaging*, vol. 37, no. 9, pp. 1970-1977, 2017.

[15] J. Wang, "Non-convex $\ell_p$ regularization for sparse reconstruction of electrical impedance tomography," in *Inverse Problems in Science and Engineering*, pp. 1-22, 2020.

[16] J. Li *et al.*, "Adaptive $\ell_p$ Regularization for Electrical Impedance Tomography," in *IEEE Sensors Journal*, vol. 19, no. 24, pp. 12297-12305, 2019.

[17] Y. Yang and J. Jia, "An image reconstruction algorithm for electrical impedance tomography using adaptive group sparsity constraint," in *IEEE Transactions on Instrumentation and Measurement*, vol. 66, no. 9, pp. 2295-2305, 2017.

[18] Y. Yang, H. Wu and J. Jia, "Image reconstruction for electrical impedance tomography using enhanced adaptive group sparsity with total variation," in *IEEE Sensors Journal*, vol. 17, no. 17, pp. 5589-5598, 2017.

[19] Y. LeCun, Y. Bengio and G. Hinton, "Deep learning," in *Nature*, vol. 521, no. 7553, pp.436-444, 2015.

[20] S. J. Hamilton and A. Hauptmann, "Deep D-Bar: Real-Time Electrical Impedance Tomography Imaging with Deep Neural Networks," in *IEEE Transactions on Medical Imaging*, vol. 37, no. 10, pp. 2367-2377, Oct. 2018, doi: 10.1109/TMI.2018.2828303.

[21] M. Capps and J. L. Mueller, "Reconstruction of Organ Boundaries with Deep Learning in the D-bar Method for Electrical Impedance Tomography," in *IEEE Transactions on Biomedical Engineering*, vol. 68, no. 3, pp. 826-833, March 2021.

[22] Z. Wei, D. Liu and X. Chen, "Dominant-current deep learning scheme for electrical impedance tomography," in *IEEE Transactions on Biomedical Engineering*, vol. 66, no. 9, pp.2546-2555, Sept. 2019, doi: 10.1109/TBME.2019.2891676.

[23] F. Li, C. Tan and F. Dong, "Electrical Resistance Tomography Image Reconstruction with Densely Connected Convolutional Neural Network," in *IEEE Transactions on Instrumentation and Measurement*, vol. 70, pp. 1-11, 2021, Art no. 4500811, doi: 10.1109/TIM.2020.3013056.

[24] Z. Chen, Y. Yang, J. Jia and P. Bagnaninchi, "Deep Learning Based Cell Imaging with Electrical Impedance Tomography," *2020 IEEE International Instrumentation and Measurement Technology Conference (I2MTC)*, Dubrovnik, Croatia, 2020, pp. 1-6, doi: 10.1109/I2MTC43012.2020.9128764.

[25] D. Hu, K. Lu and Y. Yang, "Image reconstruction for electrical impedance tomography based on spatial invariant feature maps and convolutional neural network," *2019 IEEE International Conference on Imaging Systems and Techniques (IST)*, Abu Dhabi, United Arab Emirates, 2019, pp. 1-6, doi: 10.1109/IST48021.2019.9010151.

[26] G. Liang, S. Ren, S. Zhao and F. Dong, "A Lagrange-Newton Method for EIT/UT Dual-modality Image Reconstruction," in *Sensors*, vol. 19, no. 9, p. 1966, 2019.

[27] H. Liu, S. Zhao, C. Tan and F. Dong, "A Bilateral Constrained Image Reconstruction Method Using Electrical Impedance Tomography and Ultrasonic Measurement," in *IEEE Sensors Journal*, vol. 19, no. 21, pp. 9883-9895, 1 Nov.1, 2019, doi: 10.1109/JSEN.2019.2928022.

[28] Z. Li, J. Zhang, D. Liu and J. Du, "CT Image-Guided Electrical Impedance Tomography for Medical Imaging," in *IEEE Transactions on Medical Imaging*, vol. 39, no. 6, pp. 1822-1832, June 2020, doi: 10.1109/TMI.2019.2958670.

[29] B. H. Brown and A. D. Seagar, "The Sheffield Data Collection System," in *Clinical Physics and Physiological Measurement*, vol. 8, no. 4A, p. 91, 1987.

[30] E. Somersalo, M. Cheney and D. Isaacson, "Existence and uniqueness for electrode models for electric current computed tomography," in *SIAM Journal on Applied Mathematics*, vol. 52, no. 4, pp. 1023-1040, 1992.

[31] D. B. Geselowitz, "An Application of Electrocardiographic Lead Theory to Impedance Plethysmography," in *IEEE Transactions on Biomedical Engineering*, vol. BME-18, no. 1, pp. 38-41, Jan. 1971, doi: 10.1109/TBME.1971.4502787.

[32] G. D. Finlayson, M. S. Drew and C. Lu, "Entropy minimization for shadow removal," in *International Journal of Computer Vision*, vol. 85, no. 1, pp. 35-57, 2009.

[33] R. C. Gonzales and R. E. Woods, *Digital Image Processing*, 3rd ed, Upper Saddle River, N.J: Pearson, 2008.

[34] S. Ioffe and C. Szegedy, "Batch normalization: Accelerating deep network training by reducing internal covariate shift," *International Conference on Machine Learning*, PMLR, 2015, pp. 448-456.

[35] J. Redmon and A. Farhadi, "Yolov3: An incremental improvement," 2018, *arXiv preprint arXiv:1804.02767*.

[36] K. He, X. Zhang, S. Ren and J. Sun, "Deep residual learning for image recognition," in *Proceedings of the IEEE Conference on Computer Vision and Pattern Recognition*, 2016, pp. 770-778.

[37] D. Bahdanau, K. Cho and Y. Bengio, "Neural machine translation by jointly learning to align and translate", 2014, *arXiv preprint arXiv:1409.0473*.

[38] S. Woo, J. Park, J. Y. Lee and I. S. Kweon, "Cbam: Convolutional block attention module," in *Proceedings of the European Conference on Computer Vision (ECCV)*, 2018, pp. 3-19.

[39] W. Liu, D. Anguelov, D. Erhan, C. Szegedy, S. Reed, C.Y. Fu and A.C. Berg, "Ssd: Single shot multibox detector," In *European conference on computer vision*, Oct. 2016, pp. 21-37.

[40] O. Ronneberger, P. Fischer and T. Brox, "U-net: Convolutional networks for biomedical image segmentation," In *International Conference on Medical image computing and computer-assisted intervention*, Oct. 2015, pp. 234-241.

[41] W. Liu, W. Zhou and T. Luo, "Cross-Modal Feature Integration Network for Human Eye-Fixation Prediction in RGB-D Images," in *IEEE Access*, vol. 8, pp. 202765-202773, 2020, doi: 10.1109/ACCESS.2020.3036681.

[42] Y. Yang and J. Jia, "A multi-frequency electrical impedance tomography system for real-time 2D and 3D imaging," in *Review of Scientific Instruments*, vol. 88, no. 8, p. 085110, 2017.

[43] I. Loshchilov and F. Hutter, "Fixing weight decay regularization in Adam", 2017, *arXiv preprint arXiv:1711.05101*.

[44] Y. Yang, J. Jia, N. Polydorides and H. McCann, "Effect of structured packing on EIT image reconstruction," *2014 IEEE International Conference on Imaging Systems and Techniques (IST) Proceedings*, Santorini, Greece, 2014, pp. 53-58, doi: 10.1109/IST.2014.6958445.

[45] S. Liu, J. Jia, Y. D. Zhang and Y. Yang, "Image Reconstruction in Electrical Impedance Tomography Based on Structure-Aware Sparse Bayesian Learning," in *IEEE Transactions on Medical Imaging*, vol. 37, no. 9, pp. 2090-2102, Sept. 2018, doi: 10.1109/TMI.2018.2816739.

[46] Zhou Wang, A. C. Bovik, H. R. Sheikh and E. P. Simoncelli, "Image quality assessment: from error visibility to structural similarity," in *IEEE Transactions on Image Processing*, vol. 13, no. 4, pp. 600-612, April 2004, doi: 10.1109/TIP.2003.819861.